%% file: template.tex
\documentclass{article}

\usepackage{arxiv}

\usepackage[utf8]{inputenc} 
\usepackage[T1]{fontenc}    
\usepackage{hyperref}       
\usepackage{url}            
\usepackage{booktabs}       
\usepackage{amsfonts}       
\usepackage{nicefrac}       
\usepackage{microtype}      
\usepackage{lipsum}		
\usepackage{graphicx}
\usepackage{natbib}
\usepackage{doi}

\usepackage{url}
\usepackage[table, svgnames]{xcolor}
\definecolor{lightblue}{RGB}{173,216,230}
\usepackage{adjustbox}
\newcommand{\subpart}[1]{\vspace{2pt}\noindent\textbf{#1}\noindent}
\usepackage{makecell}
\usepackage{tabu}
\usepackage{booktabs}

\usepackage{hyperref}

\title{Enabling seamless creation of annotated spaces: enhancing learning in VR environments}

\author{ 
    Maximilian Enderling\\
    Friedrich Schiller University Jena
	\And
    Jan Hombeck\\
    Friedrich Schiller University Jena
    \And
    Kai Lawonn\\
    Friedrich Schiller University Jena
}

\renewcommand{\headeright}{}
\renewcommand{\undertitle}{}
\renewcommand{\shorttitle}{Enhancing learning in VR environments}

\hypersetup{
pdftitle={A template for the arxiv style},
pdfsubject={q-bio.NC, q-bio.QM},
pdfkeywords={First keyword, Second keyword, More},
}

\begin{document}
\maketitle

\begin{abstract}
We present an approach to evaluate the efficacy of annotations in augmenting learning environments in the context of Virtual Reality. Our study extends previous work highlighting the benefits of learning based in virtual reality and introduces a method to facilitate asynchronous collaboration between educators and students. These two distinct perspectives fulfill special roles: educators aim to convey information, which learners should get familiarized. Educators are empowered to annotate static scenes on large touchscreens to supplement information. Subsequently, learners explore those annotated scenes in virtual reality.
To assess the comparative ease and usability of creating text and pen annotations, we conducted a user study with 24 participants, which assumed both roles of learners and teachers. Educators annotated static courses using provided textbook excerpts, interfacing through an 86-inch touchscreen. Learners navigated pre-designed educational courses in virtual reality to evaluate the practicality of annotations.
The utility of annotations in virtual reality garnered high ratings. Users encountered issues with the touch interface implementation and rated it with a low intuitivity. Despite this, our study underscores the significant benefits of annotations, particularly for learners. This research offers valuable insights into annotation enriched learning, emphasizing its potential to enhance students' information retention and comprehension.
\end{abstract}

\keywords{Computers and Graphics\and Virtual Reality \and Touchscreen \and Annotations \and Education}

\input{sections/01_introduction}
\input{sections/02_relatedWork}
\input{sections/03_methods}
\input{sections/04_userStudy}
\input{sections/05_results}
\input{sections/06_discussion}
\input{sections/07_conclusion}

\bibliographystyle{unsrtnat}
\bibliography{references}  






\end{document}

%% file: sections/01_introduction.tex
\section{Introduction}
The advent of Virtual Reality (VR) devices has ushered in a new era, marked by the growing popularity of VR applications across diverse domains ~\cite{vrPopularity}. In particular, the realm of education and training has witnessed a surge in innovation, with an array of applications leveraging the immersive potential of VR ~\cite{reviewVRTrainings, vrELearning}. Amidst this burgeoning landscape, our focus lies on harnessing the unique capabilities of VR for educational purposes, emphasizing a mixed environment that intertwines virtual reality with tangible interactions facilitated by large touchscreens.

The gamut of applications in VR education ranges from purely virtual worlds to hybrid setups incorporating both virtual and physical elements ~\cite{arClimbing, arAmbulance, laparoscopyInstrumentVR, weaponTrainingVR}. A common thread among these applications is the creation of virtual training environments, each adopting diverse approaches with varying levels of complexity ~\cite{vrTrainingRecording, coAssemble}. In our research, we advocate for a mixed environment where students are immersed in VR, while educators interface through a large touchscreen, similar to other existing research ~\cite{largeDisplays1, largeDisplays2}. This approach further enables the future creation of systems where educators can harness the benefits of VR in teaching, without encountering some of its problems. For instance, educators would still have the possibility to observe and engage with students in the real time, fostering a holistic understanding of student activities and enhancing assistance during the learning process.
The incorporation of large touchscreens into the educational paradigm introduces a dynamic layer to the teaching and learning process. During synchronous classes, educators gain the ability to closely observe real-world student actions, a perspective not afforded when immersed in virtual reality. This close monitoring enhances educators' grasp of students' activities, enabling more effective assistance when queries arise ~\cite{gesturalInterfaces}. Additionally, educators can maintain interaction with the physical environment, accessing various supplementary sources, including printed materials. To address the occasional need for additional information beyond the virtual training environment, educators can view a replicated representation of learners' perspectives on the touchscreen interface, leveraging its expansive display capabilities to accommodate multiple views simultaneously.

While touch interfaces are commonplace and well-studied, especially in the context of smartphones ~\cite{gesturalInterfaces}, their application in the realm of 3D controls is an active area of research ~\cite{multiTouch3D}. Our contribution extends this line of inquiry by proposing novel techniques for effective multitouch recognition on touchscreens within mixed educational spaces.
On the side of the learners, the benefits of immersive VR experiences are manifold. Research indicates that VR aids users in gaining insights across diverse fields ~\cite{vrLearningSerious}. Specialized applications targeting language learning ~\cite{vrLearningLanguage} and spatial relationship comprehension ~\cite{vrLearningSpatial} showcase the versatility of VR in enhancing the learning process ~\cite{kowalewski2018laptrain,yiannakopoulou2015virtual,vrFasterLearning}. Our work builds upon these foundations, presenting an alternative approach to generating training scenarios that balances ease of use with educational efficacy.
To achieve this balance, we empower educators to work with carefully crafted static scenes enriched with annotations ~\cite{annotations1, annotations2}. Unlike conventional approaches, our method draws inspiration from the annotation framework proposed by Marques et al.~\cite{vuforiaAnnotations}, adapting it to the unique demands of classroom and training course settings. These annotations serve as dynamic elements within the virtual space, facilitating asynchronous interactions between educators and learners.
Our empirical evaluation delves into the effectiveness of asynchronous interactions facilitated by our proposed mixed environment. The study, involving 24 participants, scrutinizes the ease of creating courses and the utility of generated annotations. This assessment focuses on the educators' effectiveness in conveying information and learners' ability to comprehend the annotated content.

In summary, our contributions encompass the design of a versatile mixed space compatible with different device types, the proposal of a method for rapidly enriching scenes with annotations, and the results of a quantitative study. Through these contributions, we aim to pave the way for the development of educational applications wherein a single teacher can interact synchronously with learners in virtual spaces.
In the subsequent sections, we delve into the intricacies of our mixed environment design, the annotation methodology, and the study's outcomes, providing valuable insights for the advancement of educational technology. This paper is accompanied by a video where each unique feature of the system we will propose is visible.
\begin{figure}
 \centering
 \includegraphics[width=\textwidth]{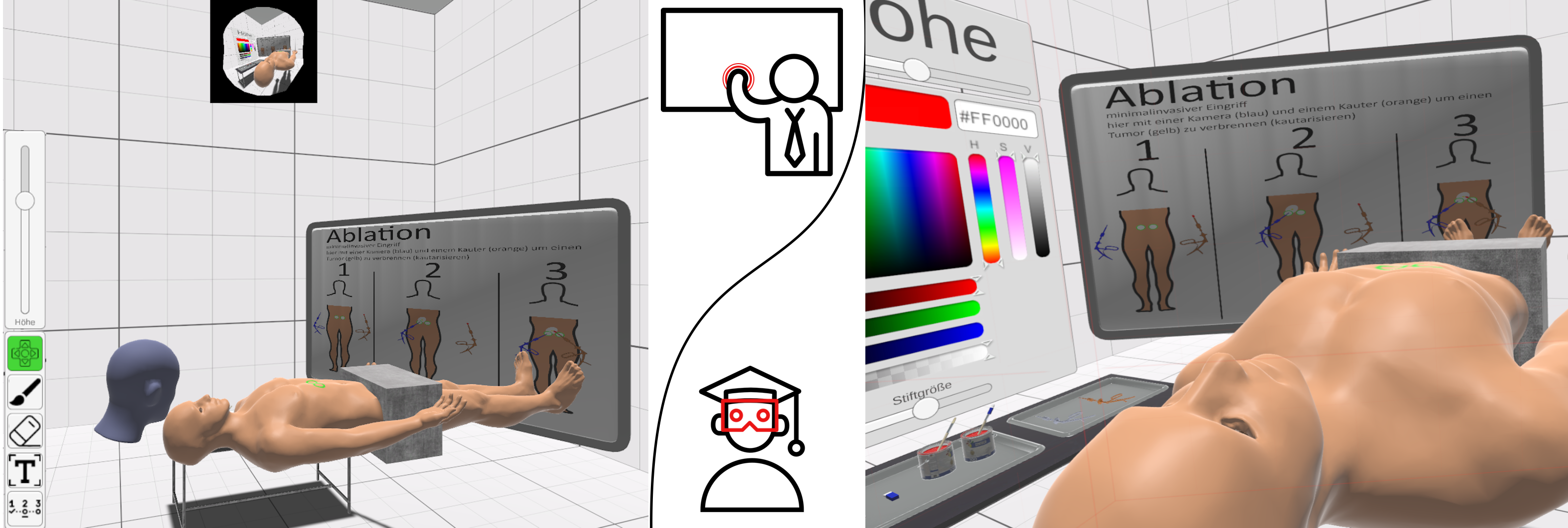}
 \caption{Views when using the system we propose in this paper. The two roles (left: educator on touchscreen, right: learner in VR) share the same environment but see different versions of user interfaces.}
 \label{fig:teaser}
\end{figure}

%% file: sections/02_relatedWork.tex
\section{Related Work}
\subsection{Touch Interfaces}
Touch interfaces utilize the flexibility of touchscreens to offer users an intuitive means of interaction with digital content, enhancing user engagement and streamlining various tasks in a wide range of applications. Current applications make use of two techniques~\cite{menuTouch, menuTouch2, touchReview}. The first technique is emulating the traditional interaction flow when using a mouse and a keyboard, resulting in buttons and menus reacting to touch. But touchscreens also enable the usage of another way of interfacing with the computer in the form of gestures~\cite{gesturalInterfaces}. This allows a more direct interaction with the environment and is shown to be less distracting for users~\cite{menuTouch2}.
There are many attempts to solve the problem of intuitive and capable movement-systems for touch interfaces. A classic attempt is to emulate physical controllers as graphical user interface (GUI). Especially, games using a first-person camera use one or two virtual joysticks~\cite{multiTouch3D}.  
The MagicCube~\cite{magicCube} one-handed technique is designed to reduce the screen area that is occluded by fingers. A GUI element is displayed, which allows movement with 5 Degrees of freedom (DOF). The GUI element shows a cube with 3 visible sides, where depending on which side a user drags from, different actions are performed such as translation, rotation, or selection of items in the environment.
Marchal et al.~\cite{multiTouch3D} summarized multiple methods and proposed a solution which is not reliant on GUI elements. It proposes a method to move back and forth and rotate left and right using one finger. when two fingers are used, a classifier determines the main component of the action. If the user rotates their fingers, the camera rotates around a point in the scene. By pinching, the camera's field of view will get regulated and by panning both fingers together, the camera can rotate both horizontally and vertically.

\subsection{Learning Systems}

\subsubsection{Touchscreens}

Touchscreens are conventional displays that are equipped with sensors that can detect whether and where the user comes into contact with them. One significant advantage is their intuitivity~\cite{touchEvaluation}: when a user want's to press a button on the screen, instead of using the mouse to move the cursor above it, they can simply tap the screen at that position. This ability results in much research into designing interfaces with them~\cite{touchM3, touchDesign}, some aimed at groups like old adults~\cite{touchInterfacesOld, touchInterfacesOld2, touchInterfacesOld3} or visually impaired individuals~\cite{touchVisualImpairment, touchVisualImpairment2, touchVisualImpairment3}.

The ubiquity of touchscreens in the form of mobile tablets and smartphones leads to many young children having access to this technology~\cite{youngChildrenTouchscreens}. There is research concerning the use of these devices for early education~\cite{touchscreenChildrenLearning, touchBenefitsDamagesKids}. They are especially beneficial to children because they have very little other contact with technology and are not accustomed to common computer interfaces like mice and keyboards.

\subsubsection{Virtual Reality}

Learning systems where learners are immersed in a VR environment received some research, especially in specialized areas where the study of real-world counterparts is expensive or difficult, but spatial information is still critical. The approach Saffo et al. ~\cite{desktopVRCombination} take is similar to the one that is described in this paper, with the contrast that they focus on interactions between VR and non-touch desktop environments. One of the fields where learning in VR is very well studied is surgical training~\cite{vrMedicalTraining, vrMedicalTraining2, vrMedicalTraining3,hombeck2024voice,laparoscopyInstrumentVR} where students are reported to learn faster~\cite{vrFasterLearning} and achieve better results~\cite{vrMedicalBetter}. Moreover, prior research has indicated that spatial and distance estimation tend to exhibit greater accuracy in VR environments when compared to desktop applications ~\cite{unityInterface2,hombeck2022distance,hombeck2019evaluation}.
Some research delves more generally into education and creates suggestions for the general architecture of educational applications. Co-assemble~\cite{coAssemble} proposes to separate learning environments into three scenarios. In single user mode, learners operate alone in an environment, having more freedom and feel less pressure. In the medium-sized setup, classes are separated into groups where learners cooperate in a shared space. Finally, the class mode is the most similar to traditional school setups; every learner is in the same space as the educator, mostly restricted to observing a live presentation. Educators may choose to allow specific learners to present things to the whole class. Each scenario has its unique benefits, so providing them all is important.

\subsection{Annotations}

Annotations provide context to existing information. Most research regarding annotations placed in 3D spaces is concerned with enabling remote collaboration or assistance~\cite{vuforiaAnnotations, annotations1, annotations2}. A common scenario is where an on-site technician requires assistance from experts. Instead of sharing individual photos, current research is trying to recreate the 3D environment of the on-site technician so that the expert can better grasp the spatial context.
Marques et al.~\cite{vuforiaAnnotations} evaluated different types of annotations regarding their usefulness to on-site technicians and remote experts. They were concerned with asynchronous assistance in three steps: first, the on-site user captures the environment and annotates it. Then the remote user inspects it, placing further annotations to detail what an intervention should look like. Thirdly, the on-site technician performs that intervention by following the provided annotations. Though they did not name all tested annotations, in their results they name the ones that were rated most useful. These annotations were (most useful first): first drawing, for its versatility, second notes, for their ability to add richer context, third notifications, to alert to information updates. Finally, they also added the possibility to add temporal sorting to other annotations, which was appreciated for its use in environments with many annotations. They noted, that editing existing annotations were an important aspect to potentially reducing workload.

%% file: sections/03_methods.tex
\section{Methods}

\subsection{Multi-Device Rendering}

Our current method requires a unique setup to render to two devices, the touchscreen and the head-mounted display (HMD), at the same time. Moreover, the view of the HMD should be included on the touchscreen in a scaled down format. We used Unity~\cite{unity} to realize this. Since the VR camera is in the same scene as the touchscreen camera, we have to specify in which order they draw.

\begin{figure}[tb]
 \centering
 \includegraphics[width=\columnwidth]{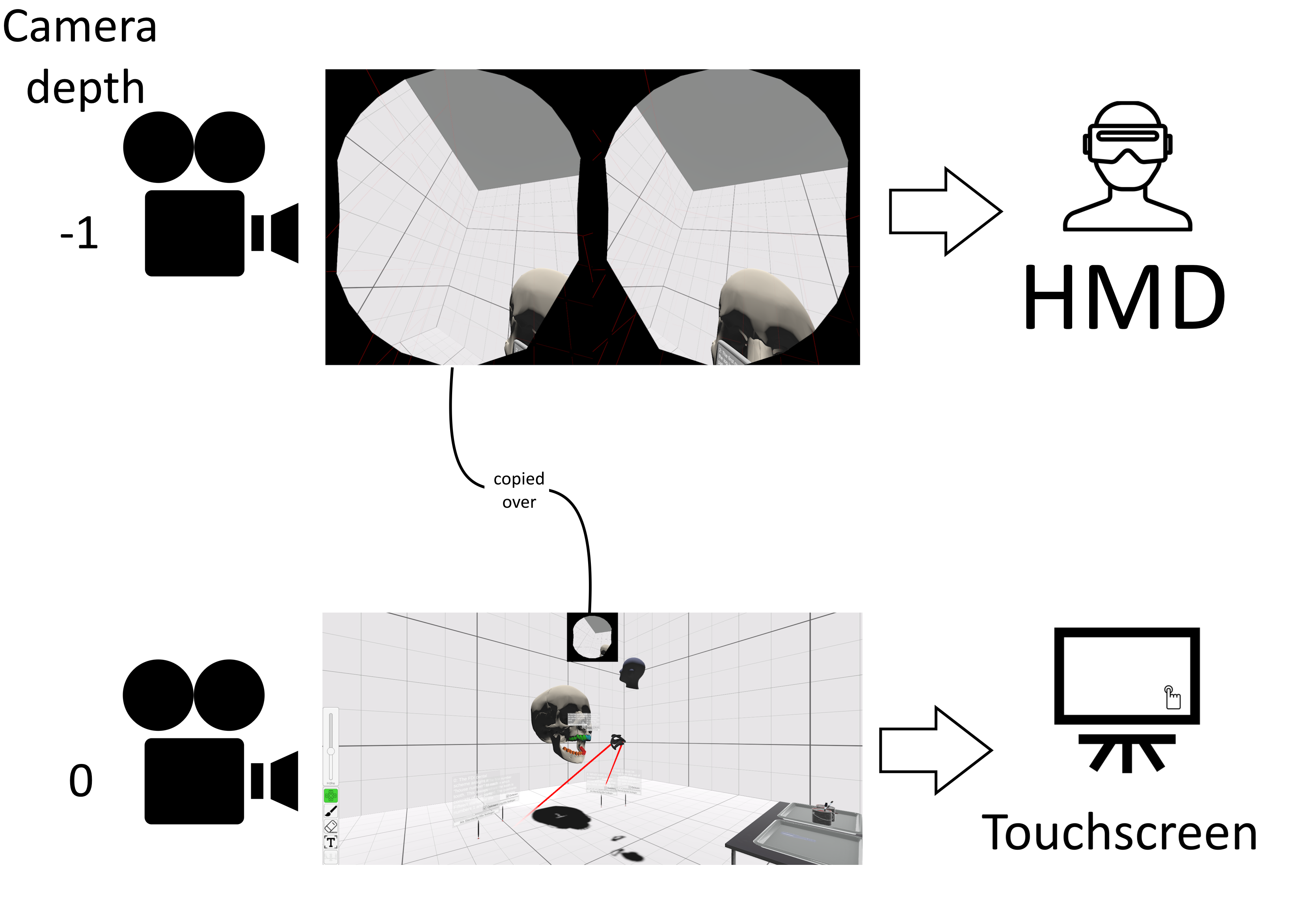}
 \caption{Camera setup designed to draw to both devices at the same time.}
 \label{fig:rendering}
\end{figure}

To control in which order the cameras draw, we have to control their depth values. First, the VR camera is rendered with a depth of -1. A script captures a view, scales it down and saves it to a texture. Then, the camera belonging to the touchscreen is rendered with a depth of 0. Inside the screen-space UI, the texture containing the VR view is included. This process is illustrated in \autoref{fig:rendering}.

\subsection{Tools}

For both the Educators using the touchscreen and the learners in VR, we provide a set of tools to enable interaction with the system. Users of the touchscreen can access them via a Toolbar, fixed at the bottom left of the screen. There, they can switch between tools, while only one tool can be active at a time.

To the learners, similar tools are provided as virtual objects, positioned in the scenes on a tray. By using the XR Interaction Toolkit\footnote{Unity XR Interaction Toolkit \url{https://docs.unity3d.com/Packages/com.unity.xr.interaction.toolkit@2.5}}, the learners could pick up objects by pointing at them while closing their hand around the controller. Opening the hand again resulted in the object being dropped. The controller is securely affixed to the back of the hand via a strap, ensuring that it remains firmly in place during operation. When objects were dropped on the floor, they are returned to their original position on the tray. The appearances of the tools in VR was chosen to be very similar to familiar tools in reality to reduce the time learners need to get accustomed to them.

\subsubsection{Movement}
\subpart{Touchscreen} To enable movement for educators, we combined multiple approaches. We opted to use a free-fly camera, similar to the Move\&Look technique~\cite{multiTouch3D}. Using a single finger to interact with the touchscreen enables the user to move towards points in space (by tapping on that point and moving their finger downward) or to move away (by moving the finger upward). Moving the finger left and right rotates the camera by applying yaw to it. Touching with two fingers and moving them, the user may rotate the camera around the yaw and pitch axis.

We introduce another way to move perpendicular to the ground in the form of a height-slider. The slider always reflects the height of the camera. By manipulating it, the user may quickly move through objects vertically. This is useful in moments, when the camera is positioned above an object, and the user wants to move below it.

\subpart{Virtual Reality} The system was designed for a Head-Mounted Display with 3D spatial tracking with 6 Degrees of Freedom. Movement in VR is thus equal to the movement in the real world, which is an intuitive solution, enabling users to move around by walking or changing their pose. To reduce the need to move or bow down, a tool was introduced in the form of another height-slider. This one is positioned above the color picker for the VR user (described below) and controls the height of the content of the scene, except for the tools. This is designed with smaller or larger users in mind.

\subsubsection{Draw, Erase and Fill}
One set of tools we supply to annotate scenes is enabling users to add colors to existing objects. This section details how users can do so. These tools were included because of their common use in other applications~\cite{vuforiaAnnotations, annotations1, annotations2}. We also provide the ability to color whole objects at once, the so-called fill tool. This is implemented by setting the albedo color of the renderer to the target value.

\subpart{Touchscreen} Educators can color objects by using the pen-tool. Once selected, they can simply swipe over the object they want to paint on to add lines. The color used is controlled by a color-picker which is displayed next to the toolbar as long as the user has the pen-tool selected. To evoke the fill tool, the user may double tap on an object to color it completely. The equivalent action using the eraser tool removes the global coloring of the object.

\subpart{Virtual Reality} For painting and erasing, we use a marker and an eraser provided by DokoDemoPainter~\cite{dokoDemo}. Additionally, a virtual brush, to change the color of an object, and an eraser on a stick, to reset the coloring, were added. A color picker, similar in form and function to the one provided on the touchscreen, is floating above the metal tray holding the tools, enabling selection of a color from a color picker restricted to the hue-value mode~\cite{hueValue}. To add the fill tool to VR, two new items were added: a brush and an eraser-brush, fulfilling their respective tasks.

\subsubsection{Text and Sequence}
Similarly to the draw tools, the text, and sequence tools were added driven by the findings of prior research that identified their utility~\cite{vuforiaAnnotations}. Our realization of text boxes can be seen at \autoref{fig:lapcamera}.

\subpart{Touchscreen} Another way to annotate scenes is to add actual text. This is more in line with traditional educational materials like textbooks. To place a text box, the user has to tap on a specific point while being in the text mode. Using a ray cast, the point in the scene that was tapped on is determined. This is chosen as the anchor for the text box. Then, a popup appears in which text may be entered using the default Windows touchscreen keyboard. To edit existing text, tapping an existing text box while the text tool is active opens a popup in which the existing text is contained. The user may delete or change the text this way. The Sequence tool works by assigning each text box an index, which dictates their order. The effects of this are detailed below in the Virtual Reality section. Tapping a text box using the sequence tool allows changing its index from the default 0.

\subpart{Virtual Reality} The learners can't create text boxes on their own. They can only interact with existing ones, marking them as ''read'' by pointing at the respective checkbox and pressing the trigger on the controller. 

The sequence is not explicitly visible to learners. Still, they benefit from the usage of this tool since they only see text boxes in a predefined order instead of all text boxes at once. The educator sets this order. Once the learner marks a text-box as 'read', the text-box which is unread and has the lowest sequence-index is displayed. If multiple text-boxes share the same index, they are all displayed at once.
Marking a text-box as read makes it smoothly shrink in size. This is done, so that other text-boxes which are near them are less likely to be occluded. While a controller is pointing at such a shrunk text-box, the text-box will regrow to normal size again.

%% file: sections/04_userStudy.tex
\section{User Study}
\subsection{Study Design}

This section aims to provide a comprehensive overview of the methodology employed to investigate the impact of different annotations on learning and teaching in touchscreen and VR environments. Participants, totaling 24, engaged in a structured study where they perform tasks as both educators and learners. Upon entering the designated study room, participants were presented with an information sheet outlining the study's context and their rights. The study was conducted on a voluntary basis, and participants were free to withdraw at any point without consequences. Upon consenting to participate, individuals completed a form and a pre-study assessment gauging their comfort level using the widely recognized Simulator Sickness Questionnaire ~\cite{ssqOriginal, ssqFix}. An introduction to the tools used in the study followed, including touch movement and VR interactions.

The main study was structured into two sections, with participants alternating between the roles of educators and learners. The decision to let all participants perform both roles was made to enable comparisons between the usability and usefulness of all tools and annotations, even though the roles would rarely be shared in real-life scenarios. Each section comprised three subtasks, structured in such a way that every participant would be exposed to all proposed annotation tools and mechanisms. To support assessment of each tool and mechanism individually, the study was designed to minimize the amount of new tools the participants were given at once, instead only introducing them one-by-one or alongside their related tools. 
Each subtask was covering a single topic, unrelated to other subtasks in this study. The topics were chosen such that participants were unlikely to have more than surface-level knowledge before participating in the study to emulate learning environments where our tools would be combined with unfamiliar information. Still, the subtasks' subjects shouldn't be too complex to be covered in the scope of this subject or require any prior knowledge that only specific professions may have.
Educators utilized drawing, fill, and eraser tools in the first subtask, followed by text and sequence tools in the second. The third subtask allowed free use of both annotation methods. The study was executed in eight variations, every participant completing all the mentioned tasks but counter-balanced to ensure each order was conducted three times. Participants, spent an average of 60–90 minutes completing the study's practical component. The study maintained consistent laboratory conditions across multiple runs by using identical hardware and software configurations. Upon completion of the practical tasks, participants filled out multiple established questionnaires~\cite{ssqOriginal, ipq, sus, daq} to provide additional insights into their experiences and perspectives. This study received ethical approval from the institution's ethics board, underscoring our commitment to maintaining the highest ethical standards in research.

\begin{figure}[tb]
 \centering
 \includegraphics[width=0.9\columnwidth]{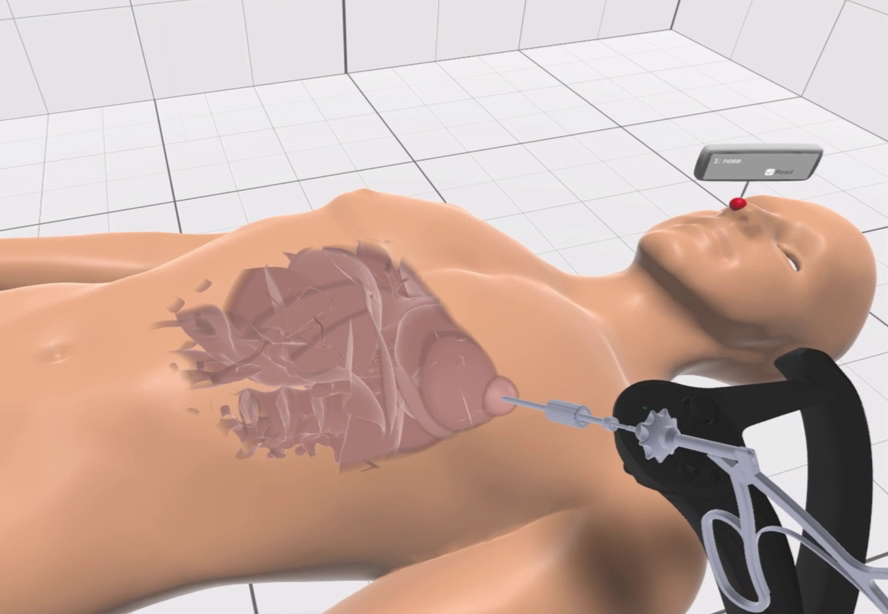}
 \caption{Illustration inserted laparoscopic camera, revealing the internal organs. An exemplary textual annotation is displayed at nose position.}
 \label{fig:lapcamera}
\end{figure}

\subsection{Hardware and Software Setup}
This section details the hardware and software that was used to implement the described method and to conduct the study. 

\subpart{Hardware} The Hardware used for this study was chosen so that both the VR and touchscreen setup could be rendered simultaneously on the same PC. Thus, the graphics card NVIDIA RTX 3090 was used together with the Intel Core i9-12900k and 32~GB of DDR5 RAM running at 4.4GHz. The VR equipment we selected was the Head-Mounted Display (HMD) Valve Index, running two LCDs at 144Hz with a per-eye resolution of 1440×1600 pixels.

\subpart{Software} The PC backing our study is running Windows 11. The methods described above were implemented in Unity 2021.3.14f1~\cite{unity}. Unity is mostly known as a game engine, but is also reasonably useful as a research tool for interface, interaction, or visualization studies~\cite{unityInterface1, unityInterface2, unityInterface3}. Unity's behavior can be customized using C\# scripts, enabling advanced functionality.

\subsection{Procedure}

The study was designed so that each participant would perform both perspectives that this method was designed for: educators and learners. This section will detail the procedure of the study.

\subpart{Introduction} The introduction was structured to introduce the participant to every tool they would use during the study. The scene prepared for this contains a human body provided by Nobutaka et al.~\cite{bodyparts3d} that was used for all depictions of anatomy throughout this study. First, the interface for educators was described by the researcher. After establishing the ways the camera could be moved, the participant was allowed to try it out. Then, the researcher demonstrated the use of the pen and the eraser on the face of a human body by drawing with blue color around one eye and red color on the lips. Part of that was erased directly afterward. Then, the researcher placed a text annotation on the nose with the text ''nose'' to demonstrate the use of the text annotation. A second one was placed on the knee. Then, the sequence tool was used to set the sequence index of the knee annotation to 2, placing it chronological behind the one on the nose.
For the second part of the introduction, the participants were equipped with the HMD. Once they reported seeing sharp images, they were allowed to move around and acclimate to the virtual environment. Once they were ready, they were told about the tools laying on a metal tray, how they could pick them up and drop them. They were asked to use all the regular annotation tools one after one. The introduction scene also contained one advanced tool, the laparoscopic camera used in the ablation scene described below, which had to be tried out.

\subsubsection{Educator} 

In the role of educators, the participants are tasked to transfer knowledge from text-book excerpts into scenes that were prepared for the study. Those excerpts were printed out so that the participants could hold them while using the touchscreen with their free hand. At the beginning of each subtask, the participants were given a small oral introduction into the specific topic and handed the excerpt to read through. Afterward, they were told which annotations they were allowed to place and shown the prepared scene. Each scene consisted of three parts which were labeled ''1.'', ''2.'' and ''3.'', later on regarded as the first, second or third part. There was no target of how much or what kind of information they were to transfer. Each subtask was finished as soon as the participant wished to do so.

In the following, each scene and the covered topic are explained in more detail. 

\subpart{Scar revision} Scar revision is the process of trying to remove or change the appearance of scars. One of the procedures to do this is to remove the scar by cutting it out and sewing the surrounding skin back together. This may be done to return flexibility when scars are near joints, or for aesthetic purposes. The excerpt handed to the participant details that and some alternative approaches to do scar revision. The three parts depict one skin piece with a scar and two identical skin pieces, where the scar is missing, and the skin has a visible hole. This scene was to be annotated only using the pen, eraser, and fill tools.

\subpart{Port} The scene covering the topic of medical ports was designed around restricting the participants to using the text and sequence tools. Medical ports are small gadgets which are implanted beneath the skin. A catheter connects them to a vein, allowing  access to the blood system of a patient without needing to puncture their veins. They are commonly used for cancer patients. The excerpt given to the participants detailed information on ports regarding their use, structure, and benefits. The first part of the scene shows a body where specific veins are visible through the skin. The most important vein for this procedure is highlighted. In the second part, the model of a port chamber is depicted. Finally, in the third, a human with a fully implanted port including the catheter is shown, with relevant parts visible through the skin.

\subpart{Anaphylaxis} The final task in the role of the educator is covering the topic of severe allergic reactions, their diagnosis and how to treat them in the field. The most commonly used drug for this is adrenaline, which gets injected into large muscles of the affected person. The excerpt covers this topic in depth, including symptoms and treatment in the field. The first part of the scene shows a human body, the second an enlarged version of an EpiPen and finally the third, the EpiPen stuck to the upper thigh. This time, the participants were allowed to make full use of all the tools.

\subsubsection{Learners}

In the part of the study where participants were exploring the systems designed for learners, they were wearing the supplied HMD. For this, they had to go through the courses described below. 

\subpart{Ablation} In this task, the learners were standing in a scene with a male body and a whiteboard. The whiteboard displays a simplified endoscopic ablation to remove a tumor. For this, two tools with needles, one with a camera and one with a hot tip, are inserted into the body. The green circles that are depicted on the whiteboard are drawn onto the body to aid the participant in locating the tumor. The participant has to insert the needle with the camera into the body. Once inserted, organs near the camera can be seen through the skin. The participant then has to search for the bright color-coded tumor and touch it with the laparoscopic tool they have holding in their other hand. Once that's done, they have finished this task. The annotations that were used to create this scene were like those created by the pen tool.

\subpart{Digestion} This scene holds the organs that are part of the human digestive tract. Each of the important parts had a text box associated with it, totaling 11 annotations. The sequence tool was used to make the text boxes appear in the order in which food would pass through them. Participants were tasked to read through every text box. For this, they also had to move around because the text boxes were occluded by the organs from some view points.

\subpart{Teeth inspection} The final task in the role of learners was concerned with teeth. They were presented with a skill with an open mouth. The teeth are colored according to their quadrant. The participants learned the Federation Dentaire Internationale (FDI) dental notation~\cite{isoTeeth}, which gives each tooth a simple ID from text boxes. Then they were tasked to mark a specific tooth with by coloring it yellow. Lastly, they should use a special tool to remove a tooth given by its ID by touching it. This scene combined all annotations we examined in this paper.

\subsection{Participants}

Of our 24 participants, 15 were male and 8 were female. The youngest participant was 23, the oldest 38 with a median age of 25. All participants had at least a High school diploma, with 4 having or aspiring to have a bachelors degree, 10 a masters-level degree and 2 PhDs. Out of the 6 people not currently being a student, 2 work as software developers and 3 as researchers.

The previous experience in the areas our research is touching is very diverse. 11 of our participants have none to very little experience with 3D environments like simulations or video games. Of the other 13, one has 3–4 years of experience, while the rest have 5+ years.
Our participants were less experienced in general with VR, with fifteen having reported no true experience, three having less than 1 year, three in the range of 1–2 years, one having 3-4 and only two people having at least 5 years of experience. Nobody reported using VR regularly. 
Except for 2 people, everyone was interacting with PCs every day. Nine participants reported playing video games every day, and six others at least a few times a month.

\subsection{Measures}
\label{sec:measures}
\subsubsection{Objective Measures}

During the tasks, we created a log where the most important events are appended with a timestamp. An entry was added each time:

\begin{itemize}
    \item a Unity scene changes, which occurred when switching to the next task,
    \item the learner finishes a task,
    \item the educator switches to a different interaction mode,
    \item the learner is using a VR controller to pick something up,
    \item or the educator is interacting with the height slider (at most once per second).
\end{itemize}

These timestamped events allow us to reconstruct how long each participant took for which tasks, what tools they used and how long they used them. To capture the drawn annotations that were created, we created photos of the scene from multiple angles when a task was finished or before a new scene was entered.

\subsubsection{Subjective Measures}
This study utilizes multiple questionnaires to measure how users reacted to our presented tools. First, to measure symptoms of sickness, the widely used \textbf{Simulator Sickness Questionnaire (SSQ)}~\cite{ssqOriginal} is included. The SSQ asks the user to rate their sickness by ranking their current feeling regarding 16 different criteria (e.g., headache, fatigue, nausea, or dizziness) on a scale of 1 to 4. Using that, 3 scales can be derived that relate to nausea (N), oculomotor disturbance (O) and disorientation (D). Kennedy et al.~\cite{ssqOriginal} report multiple thresholds with $>20$ being the most severe one, relating to a bad simulator~\cite{ssqFix}. Though, as noted by Bimberg et al.~\cite{ssqFix}, we applied the corrected formula for the final score and used the common approach of using a pre- and post-study questionnaire. We wanted to assess how the SSQ scores differed between the usage of touchscreens and the HMD, so we separated the post-study SSQ into two parts, where participants had to rank their feelings for each of the device types.
Even when somebody does not actually experience symptoms of simulator sickness, they still may feel other types of discomfort regarding specific tools. To quantize this, we employ the \textbf{Device Assessment Questionnaire (DAQ)}~\cite{daq} in which each volunteer had to rate the required force, smoothness, mental and physical effort as well as various bodily fatigues (totaling thirteen properties) when using each of the eleven tools. This resulted in 143 ratings the users had to perform. The scale of the DAQ goes from 1 to 5, while the interpretation was inconsistent between each of the questions. For some questions, 3 is the ideal case, where for some it is 5 as seen in the header of \autoref{tab:daq}.
We also want to measure presence, as that is a big factor for learning experiences ~\cite{presenceLearning}. This was gauged by adding the \textbf{Igroup Presence Questionnaire (IPQ)}~\cite{ipq} to our post-study questionnaire. The fourteen questions included try to measure how much the simulated world was recognized and experienced as reality. They also ask the participant to reflect on how much of the real world was still perceived. The questionnaire presents statements and asks the responded to rate their agreement from 0 through 7. The questions were again separated between touchscreen and VR.

To assess the usability of our tools more directly, users had to answer the \textbf{System Usability Scale (SUS)}~\cite{sus}. Each of the 10 statements contained tried to address different important factors for real-world usability, which users had to rate their agreement with. To consider our tools independently, each of the 10 statements had to be rated for each of the 11 tools, resulting in 110 ratings. The rating was conducted by assigning each of the statements a score between 1 and 5, where 1 represents minimal agreement and 5 maximum agreement with the statement. The usability can then be calculated according to SUS and falls between 0 and 100.

On the last page, users had to state which of the 6 subtasks was their favorite. Finally, users had to evaluate the perceived usefulness of the annotations, regardless of their current implementation.

%% file: sections/05_results.tex
\section{Results}
\begin{figure}[tb]
    \centering
    \begin{minipage}{0.49\columnwidth}
        \centering
        \includegraphics[width=\linewidth]{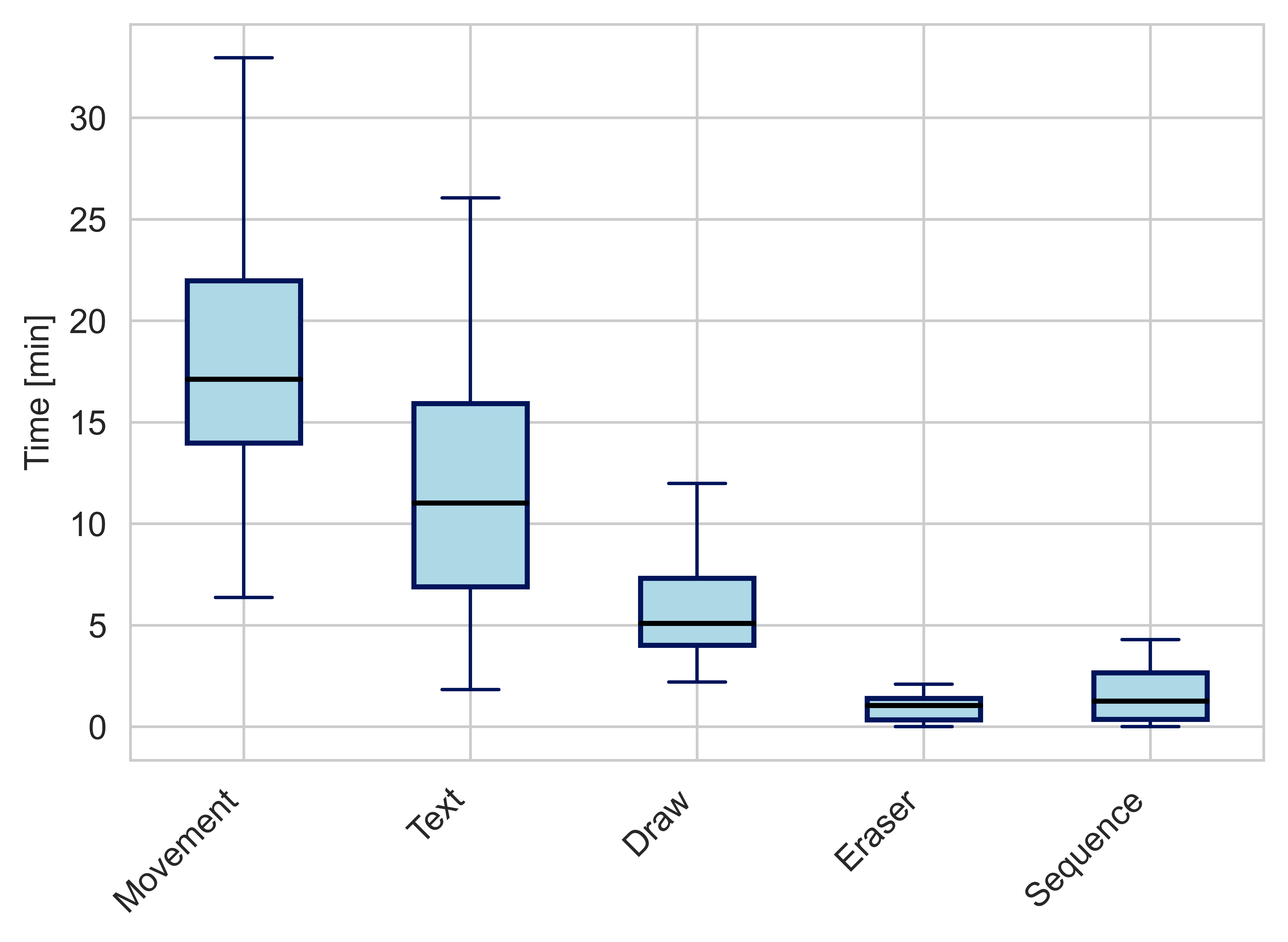}
        \caption{The time in minutes the participants used each tool on the touchscreen.}
        \label{fig:touchToolUsage}
    \end{minipage}
    \hfill
    \begin{minipage}{0.49\columnwidth}
        \centering
        \includegraphics[width=\linewidth]{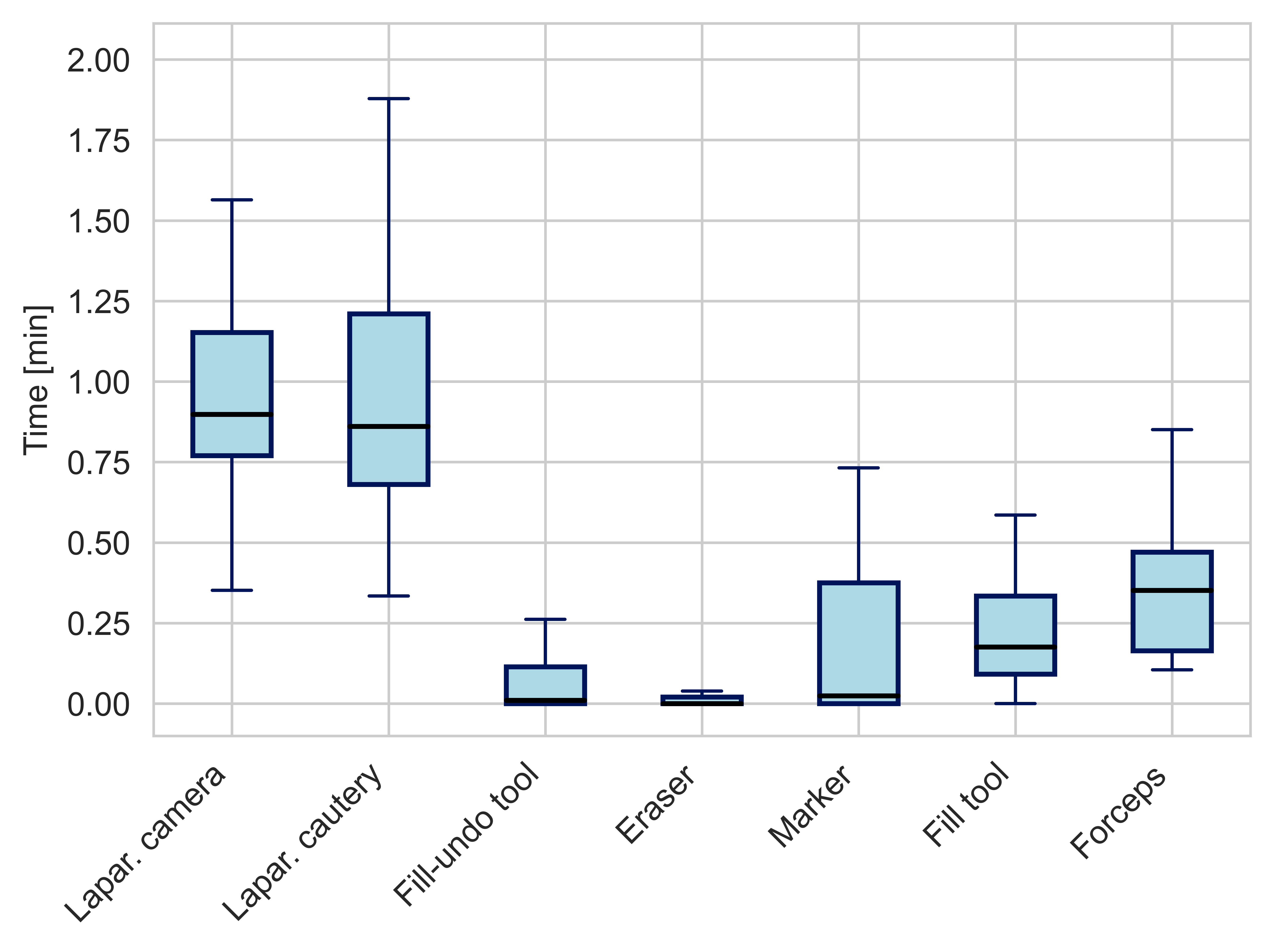}
        \caption{The time in minutes the participants used each tool in VR.}
        \label{fig:vrToolUsage}
    \end{minipage}
\end{figure}

\begin{figure}[t]
    \centering
    \begin{minipage}{0.49\columnwidth}
        \centering
        \includegraphics[width=\linewidth]{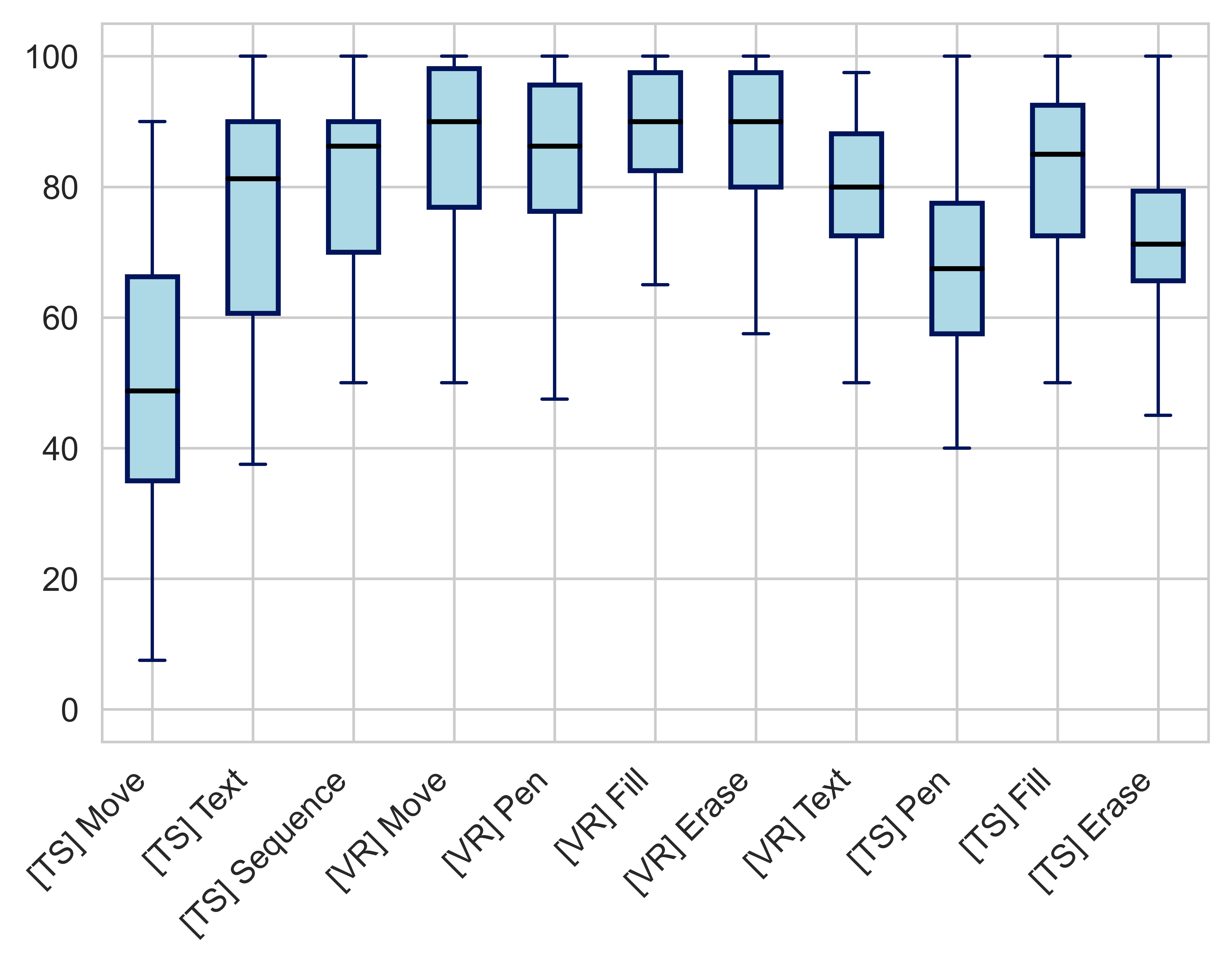}
        \caption{Box plot of the SUS~\cite{sus} usability scores for the different tools. TS = Touchscreen.}
        \label{fig:usability}
    \end{minipage}
    \hfill
    \begin{minipage}{0.49\columnwidth}
        \centering
        \includegraphics[width=\linewidth]{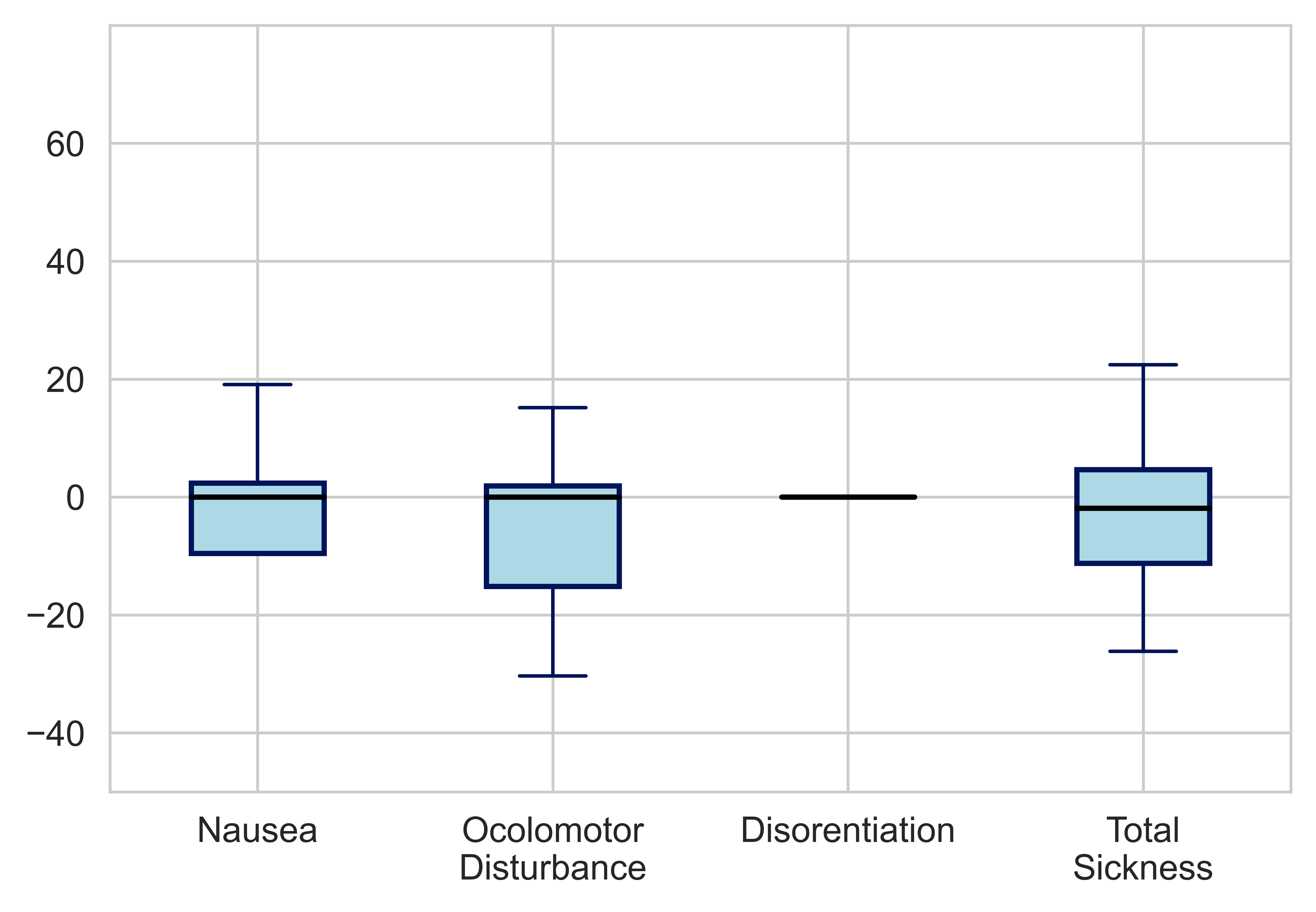}
        \vspace{1em}
        \caption{Box plot of the recorded SSQ~\cite{ssqOriginal, ssqFix} scores for the touchscreen.}
        \label{fig:touchSickness}
    \end{minipage}
\end{figure}

\subsection{Objective Measures}

We recorded the time the users spent during each scene. It is important to note that, in most cases, the time that the user needed to complete the pre-study questionnaire is included in the initial scene length. Moreover, the clock was running while the users read the excerpt that was given to them for the scenes where they had to act as an educator. This was performed because multiple participants did not read the whole text at the beginning, which would have enabled simple measurement of that duration. The average time spent on the learner tasks, Ablation ($1.7\pm{}0.7 min$), Digestion ($3.4\pm{}1.4 min$), Teeth inspection ($3.2\pm{}1.3 min$) were significantly shorter than the ones spent on annotating scenes as an educator, which were Scar revision ($11.4\pm{}4.3 min$), Port ($14.4\pm{}6.7 min$) and Anaphylaxis ($11.5\pm{}5.1 min$). There were 3 guided scenes, namely all 3 learner-tasks, where users completed tasks without applying much creativity. Still, the coefficient of variation ($standard deviation / mean$) normalizes the standard deviation and is around $0.37$ to $0.46$ on all scene lengths.
On average, the participants spent 37.4 minutes in the educator's role and only 8.3 minutes as learners. This, combined with the average 18 minutes which the introduction and pre-study assessment took, participants spent on average 63 minutes and 39 minutes until the end of the practical part of the study. Afterward, they were still required to complete the post-study questionnaire, which was not time-limited.
As our study log included details regarding which tool was switched, we could reconstruct the time each tool was active. This is not necessarily the amount of time the tools were in use because idling would not be excluded. The mean average use time for the tools available on the touch screen were: Movement ($1055.9\pm{}383.0s$), Text ($667.0\pm{}384.8s$), Draw ($339.2\pm{}156.2s$), Eraser ($67.9\pm{}57.9s$), Sequence ($124.7\pm{}98.7s$). In total, the users spent 46.8\% of their time as an educator using the movement mechanics. A detailed distribution is depicted in \autoref{fig:touchToolUsage}.

The tools held in VR were also tracked and can be observed in \autoref{fig:vrToolUsage}. While the timer ran while the participants were just holding the tools, research personnel observed very little unnecessary retention of tools because users placed an emphasis on dropping tools as soon as they didn't need them anymore. Some users ended up automatically grabbing their controllers, which lead to them picking up tools without then noticing. Mean usage times for the tools: laparoscopic camera ($58.5\pm{}23.0s$), laparoscopic cautery ($54.5\pm{}22.9s$),  eraser ($2.5\pm{}7.9s$), marker ($14.3\pm{}19.6s$), fill tool ($18.4\pm{}23.2s$), fill-undo tool ($6.0\pm{}11.6s$), Forceps ($26.8\pm{}25.7s$).

\begin{figure}[t]
    \centering
    \begin{minipage}[t]{0.49\columnwidth}
        \centering
        \raisebox{0pt}[\dimexpr\height+1\baselineskip\relax]{
            \includegraphics[width=\linewidth]{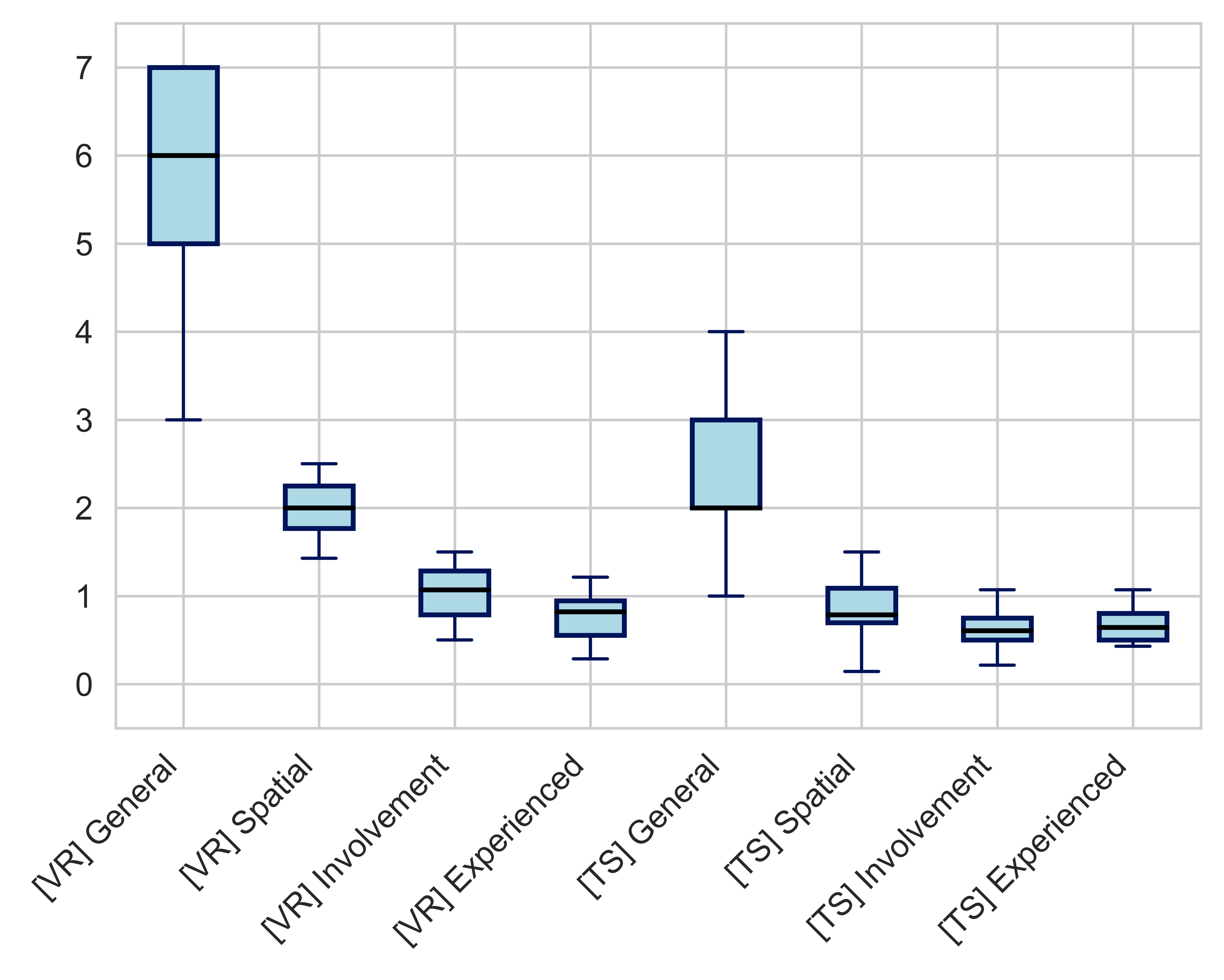}
        }
        \caption{The IPQ presence scores~\cite{ipq}, calculated from the responses given by the participants. TS = Touchscreen.}
        \label{fig:presence}
    \end{minipage}
    \hfill
    \begin{minipage}[t]{0.49\columnwidth}
        \centering
        \raisebox{26pt}[\dimexpr\height+1\baselineskip\relax]{
            \includegraphics[width=\linewidth]{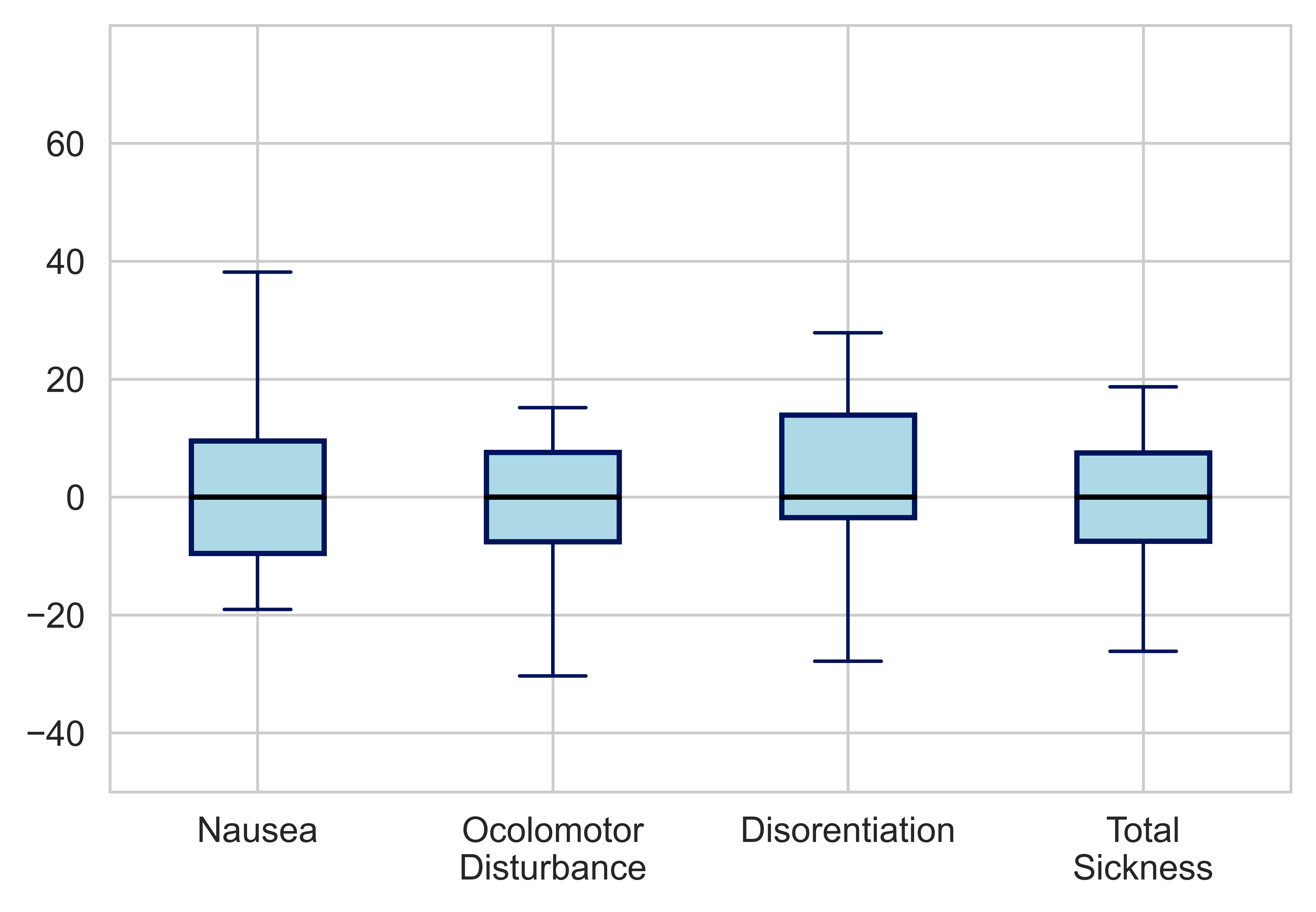}
        }
        \caption{Box plot for the recorded SSQ~\cite{ssqOriginal, ssqFix} scores for VR.}
        \label{fig:vrSickness}
    \end{minipage}
\end{figure}

\begin{table*}[b]
    \centering
        \caption{Mean ± standard deviation for the results of the DAQ~\cite{daq}. Cell color intensity indicates the magnitude of difference to the optimum. opt.=optimum; fat. = fatigue}
    \label{tab:daq}
     \begin{adjustbox}{width=\textwidth}
    \begin{tabular}{rlllllllllllll}
Tool & \makecell{Smoothness\\opt.:5}  & \makecell{Accuracy\\opt.:5}  & \makecell{Comfort\\opt.:5}  & \makecell{Usability\\opt.:5} & \makecell{Finger fat.\\opt.:1} & \makecell{Wrist fat.\\opt.:1} & \makecell{Arm fat.\\opt.:1} & \makecell{Shoulder fat.\\opt.:1} & \makecell{Neck fat.\\opt.:1} & \makecell{Mental effort\\opt.:3}  & \makecell{Physical effort\\opt.:3} & \makecell{Force required \\opt.:3}& \makecell{Speed\\opt.:3} \\
\toprule
Touch Move & 2.75±1.22\cellcolor{lightblue!56} & 2.75±1.45\cellcolor{lightblue!56} & 2.50±1.06\cellcolor{lightblue!62} & 2.58±1.10\cellcolor{lightblue!60} & 2.25±1.59\cellcolor{lightblue!31} & 1.42±0.97\cellcolor{lightblue!10} & 1.83±1.27\cellcolor{lightblue!20} & 1.62±1.01\cellcolor{lightblue!15} & 1.21±0.51\cellcolor{lightblue!5} & 3.75±0.74\cellcolor{lightblue!37} & 3.42±0.88\cellcolor{lightblue!20} & 3.29±0.62\cellcolor{lightblue!14} & 3.17±0.70\cellcolor{lightblue!8} \\
Touch Pen & 2.43±1.44\cellcolor{lightblue!64} & 2.70±1.52\cellcolor{lightblue!57} & 2.87±1.06\cellcolor{lightblue!53} & 3.00±1.17\cellcolor{lightblue!50} & 1.78±1.09\cellcolor{lightblue!19} & 1.26±0.62\cellcolor{lightblue!6} & 1.43±0.66\cellcolor{lightblue!10} & 1.39±0.72\cellcolor{lightblue!9} & 1.17±0.49\cellcolor{lightblue!4} & 3.09±0.29\cellcolor{lightblue!4} & 3.13±0.76\cellcolor{lightblue!6} & 3.22±0.90\cellcolor{lightblue!10} & 3.26±0.62\cellcolor{lightblue!13} \\
Touch Fill & 4.00±1.15\cellcolor{lightblue!25} & 4.21±1.08\cellcolor{lightblue!19} & 3.89±1.20\cellcolor{lightblue!27} & 4.26±0.87\cellcolor{lightblue!18} & 1.26±0.56\cellcolor{lightblue!6} & 1.16±0.50\cellcolor{lightblue!3} & 1.21±0.54\cellcolor{lightblue!5} & 1.21±0.54\cellcolor{lightblue!5} & 1.16±0.50\cellcolor{lightblue!3} & 2.84±0.37\cellcolor{lightblue!7} & 2.79±0.42\cellcolor{lightblue!10} & 3.05±0.52\cellcolor{lightblue!2} & 2.79±0.54\cellcolor{lightblue!10} \\
Touch Erase & 2.32±1.17\cellcolor{lightblue!67} & 2.82±1.53\cellcolor{lightblue!54} & 3.00±1.07\cellcolor{lightblue!50} & 3.45±1.10\cellcolor{lightblue!38} & 1.59±0.96\cellcolor{lightblue!14} & 1.41±0.96\cellcolor{lightblue!10} & 1.73±1.03\cellcolor{lightblue!18} & 1.41±0.67\cellcolor{lightblue!10} & 1.18±0.50\cellcolor{lightblue!4} & 3.09±0.29\cellcolor{lightblue!4} & 3.18±0.80\cellcolor{lightblue!9} & 3.41±0.67\cellcolor{lightblue!20} & 3.32±0.65\cellcolor{lightblue!15} \\
Touch Text & 3.46±1.28\cellcolor{lightblue!38} & 3.92±1.38\cellcolor{lightblue!27} & 3.38±1.28\cellcolor{lightblue!40} & 3.79±1.35\cellcolor{lightblue!30} & 1.67±1.09\cellcolor{lightblue!16} & 1.17±0.56\cellcolor{lightblue!4} & 1.33±0.56\cellcolor{lightblue!8} & 1.25±0.53\cellcolor{lightblue!6} & 1.29±0.62\cellcolor{lightblue!7} & 3.25±0.68\cellcolor{lightblue!12} & 3.25±0.79\cellcolor{lightblue!12} & 3.04±0.36\cellcolor{lightblue!2} & 3.25±0.68\cellcolor{lightblue!12} \\
Touch Sequence & 3.86±1.13\cellcolor{lightblue!28} & 4.23±1.02\cellcolor{lightblue!19} & 3.68±1.29\cellcolor{lightblue!32} & 4.27±0.94\cellcolor{lightblue!18} & 1.32±0.65\cellcolor{lightblue!7} & 1.14±0.47\cellcolor{lightblue!3} & 1.23±0.53\cellcolor{lightblue!5} & 1.18±0.50\cellcolor{lightblue!4} & 1.18±0.50\cellcolor{lightblue!4} & 3.14±0.77\cellcolor{lightblue!6} & 3.05±0.58\cellcolor{lightblue!2} & 3.05±0.38\cellcolor{lightblue!2} & 3.09±0.43\cellcolor{lightblue!4} \\
\midrule
VR Move & 4.13±1.25\cellcolor{lightblue!21} & 4.52±0.99\cellcolor{lightblue!11} & 4.09±1.24\cellcolor{lightblue!22} & 4.52±0.99\cellcolor{lightblue!11} & 1.13±0.46\cellcolor{lightblue!3} & 1.09±0.42\cellcolor{lightblue!2} & 1.22±0.60\cellcolor{lightblue!5} & 1.17±0.49\cellcolor{lightblue!4} & 1.43±0.66\cellcolor{lightblue!10} & 2.96±0.37\cellcolor{lightblue!2} & 2.87±0.46\cellcolor{lightblue!6} & 3.00±0.52\cellcolor{lightblue!0} & 2.87±0.46\cellcolor{lightblue!6} \\
VR Pen & 3.75±1.36\cellcolor{lightblue!31} & 4.21±1.14\cellcolor{lightblue!19} & 3.83±1.34\cellcolor{lightblue!29} & 4.29±0.91\cellcolor{lightblue!17} & 1.38±0.71\cellcolor{lightblue!9} & 1.29±0.75\cellcolor{lightblue!7} & 1.42±0.72\cellcolor{lightblue!10} & 1.29±0.62\cellcolor{lightblue!7} & 1.29±0.62\cellcolor{lightblue!7} & 2.96±0.46\cellcolor{lightblue!2} & 2.88±0.45\cellcolor{lightblue!6} & 3.08±0.78\cellcolor{lightblue!4} & 2.92±0.50\cellcolor{lightblue!4} \\
VR Fill & 4.27±0.94\cellcolor{lightblue!18} & 4.55±0.86\cellcolor{lightblue!11} & 4.00±1.23\cellcolor{lightblue!25} & 4.64±0.49\cellcolor{lightblue!9} & 1.23±0.61\cellcolor{lightblue!5} & 1.18±0.50\cellcolor{lightblue!4} & 1.14±0.47\cellcolor{lightblue!3} & 1.18±0.50\cellcolor{lightblue!4} & 1.27±0.63\cellcolor{lightblue!6} & 2.95±0.38\cellcolor{lightblue!2} & 2.82±0.39\cellcolor{lightblue!9} & 3.05±0.49\cellcolor{lightblue!2} & 2.91±0.29\cellcolor{lightblue!4} \\
VR Erase & 3.77±1.19\cellcolor{lightblue!30} & 4.00±1.31\cellcolor{lightblue!25} & 3.82±1.26\cellcolor{lightblue!29} & 4.27±0.88\cellcolor{lightblue!18} & 1.36±0.73\cellcolor{lightblue!9} & 1.45±0.96\cellcolor{lightblue!11} & 1.45±0.86\cellcolor{lightblue!11} & 1.32±0.65\cellcolor{lightblue!7} & 1.36±0.66\cellcolor{lightblue!9} & 2.95±0.38\cellcolor{lightblue!2} & 3.00±0.62\cellcolor{lightblue!0} & 3.00±0.62\cellcolor{lightblue!0} & 2.95±0.58\cellcolor{lightblue!2} \\
VR Text& 3.50±1.14\cellcolor{lightblue!37} & 3.55±0.96\cellcolor{lightblue!36} & 3.77±1.19\cellcolor{lightblue!30} & 3.95±0.72\cellcolor{lightblue!26} & 1.23±0.61\cellcolor{lightblue!5} & 1.09±0.43\cellcolor{lightblue!2} & 1.27±0.63\cellcolor{lightblue!6} & 1.27±0.63\cellcolor{lightblue!6} & 1.41±0.67\cellcolor{lightblue!10} & 3.14±0.64\cellcolor{lightblue!6} & 3.05±0.58\cellcolor{lightblue!2} & 3.27±0.63\cellcolor{lightblue!13} & 3.09±0.43\cellcolor{lightblue!4} \\
\bottomrule
\end{tabular}
\end{adjustbox}
\end{table*}

\subsection{Subjective Measures}

The SSQ scores for VR in \autoref{fig:vrSickness} and the touchscreen in \autoref{fig:touchSickness} are represented as box plots. Here, higher scores are undesirable and values above 20 are typically interpreted as bad~\cite{ssqOriginal}. It can be observed that the median for all sickness values lies at 0. The box contains every value that falls in the second and third quartile. Thus, every point above the upper end of the box falls within the worst 25\% of reported values. The whiskers above and below each pox extend to include the minimum and maximum values. Anything outside the whiskers was detected as an outlier. It should be noted that values below 0 are achieved because participants reported more sickness pre-study when compared to post-study reporting. This should not be interpreted that our simulator is helpful in reducing sickness, but rather the sickness diminishing over time~\cite{ssqFix}.
The results of the DAQ are reported in \autoref{tab:daq}. Since the optimal values differ, we colored each cell to indicate how unfavorable the judgements of the participants were. For the last 4 columns, the reported value can either be too high (above 3), or it can be too low (below 3). This allows the participants, e.g., the option to describe the mental effort as too little, which could lead to slips, or too high, which tires users more than necessary.
We applied the IPQ to assess the experienced presence of our volunteers. The users gave a list of statements a score from 0 to 7. Different statements pertain to distinct categories within the IPQ. The resulting scores can be found in \autoref{fig:presence}. While presence is higher in VR than when using the touchscreen, the general presence is the highest for both.
The last questionnaire provided the users with a direct opportunity to rate the tools. The resulting SUS scores which are depicted in \autoref{fig:usability} range from 0 to 100. In general, VR tools were rated the best.
Users also had to choose their favorite task. From the 24 participants, 14 chose the teeth inspection, 5 the digestion and 4 the ablation task. Only one participant liked any of the educators tasks the best, which was the port in this case. At last, participants rated each of the annotations regarding their utility, as seen in \autoref{fig:annotationsUse}.

%% file: sections/06_discussion.tex
\section{Discussion}

\subsection{Objective Measures}

We recorded the duration each participant spent in the different scenarios, which revealed greatly varying timescales, with the teachers' tasks taking around 3–5 times longer than the learners' tasks. While they not directly comparable, they were designed to be of approximately equal complexity. When reviewing the coefficients of variation, the time needed to plan lessons appears to be equally predictable, indicating that while planning a lesson for learners to complete, the duration most students will take is not too variable. The quick completion times for learners also hint at good comprehensibility of the annotations. None of the participants were observed to be confused for more than a few seconds while being in the learners' role. The educators' role proved more challenging, which is reflected in the higher duration per task. Reading through textbook material of a different specialty proved to be challenging, as participants reported.

As discussed in the results, educators spend 46.8\% percent of their time using the movement tool. Ideally, this would be used only very briefly between the use of other tools. This statistic indicates that the movement techniques utilized in this paper can still be optimized. Next in line is the text tool, which is explained by the fact that entering text takes time. It should also be considered, that only a few of the participants were already familiar with the topics covered. Participants frequently consulted the textbook excerpt again while already in text mode. Tool use in VR was very limited. This was intentional, since even while learners had access to the tools, they mostly observed already existing annotations by reading through text.

\subsection{Subjective Measures}

The SSQ provides us with measurements regarding simulator sickness. Only a few of our test subjects experienced simulator sickness. The median value for every scale in the SSQ is observed to be 0. Values below 0 should be considered as 0 when interpreting the SSQ~\cite{ssqFix}. Though, the disorientation subscale shows that more than 25\% of our test subjects at least some disorientation experienced. As was to be expected, the touchscreen was not at all sickening, according to the SSQ. This gives us assurance, that the sickness experienced using VR was not induced from problems with the virtual environment we designed, but rather from the use of VR. Only 2 out of our 24 participants experienced a total sickness score higher than 20, both of which reported having little to no prior experience with VR, with which they could have compared their experiences during the study. Their sickness could be explained by the human factor ~\cite{vrSicknessReasons}.

In \autoref{tab:daq} we observe a clear divide between the VR and touchscreen tools. Users clearly favored the VR tools, while the three touch tools to move, paint and erase seem to have issues in their current implementation. The move tool was reported to cause finger fatigue and require too much mental and physical effort. Most of the touch tools were reported to cause slight arm and shoulder fatigue, which is probably caused by the fact that gestures require you to move your whole arms. The effort, force, and speed were considered excellent for the VR tools.

To measure presence, we consider the IPQ scores. When comparing each of the scores between VR and the touchscreen, VR provides more presence. With a median of 6, the general presence reported for VR is very high. As discussed earlier, this was desired since it is shown to aid in learning~\cite{presenceLearning}. When using the touchscreen, participants experienced less presence. This ensures that they are still aware of their surroundings and, later on, aware of the learners in their class.

The SUS scores once again gave participants the option to share their experiences with the tools. The divide between VR and touchscreen tools is a bit less clear now. Users communicated orally during their study the reasons for dissatisfaction. Regarding the movement on the touchscreen, many had issues with its complexity. Participants used the provided height-slider sporadically. It is possible, that after a longer acclimatization period, they would adapt to it better. Many expressed wishes to rotate the camera around a point in space, teleport to specific points, use pinch to move forward or add GUI-based directional pads. Compared to the simplicity of moving in VR, this underlines the need for more research into 3D touch movement controls. Issues with the touch pen and erase mostly stem from inconsistent pen sizes and problems with the library causing the pen to draw through body parts. While some volunteers disliked having to use the touchscreen for keyboard input when writing text, others found it dynamic and adapted the size and layout of the keys to their needs. Based on this situation, we recommend providing a physical wireless keyboard when users wish to enter more than a few characters. When participants were in the role of the learner, they sometimes observed text boxes to be occluded by other parts of the scene but could mitigate this issue by moving around, which caused the text box to rotate. Still, other solutions like text boxes that draw on top of other geometry, possibly only when pointed at, come to mind and could be explored in future research.

\begin{figure}[tb]
 \centering
 \includegraphics[width=0.8\columnwidth]{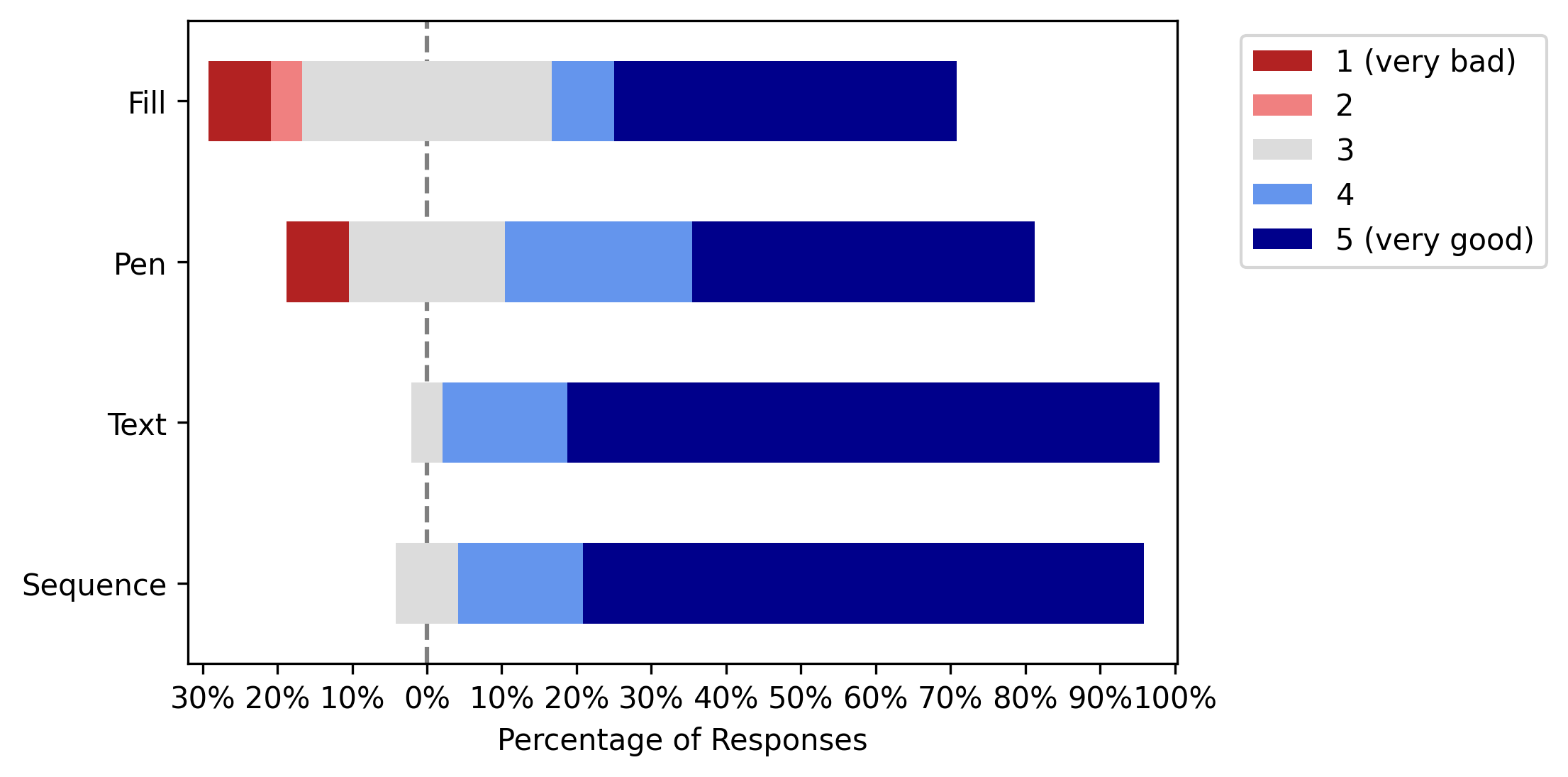}
 \caption{Percentage of user reports for each of the utility ratings of specific annotations in the context of learning.}
 \label{fig:annotationsUse}
\end{figure}

Users enjoyed experiencing the courses in the role of the learner. This was reflected in the choice of their favorite task, where 23 out of 24 chose one of the learner tasks. The study was concluded with \autoref{fig:annotationsUse}, which demonstrates the perceived utility of the provided annotation techniques for learning contexts. For each of the annotations, at least a simple majority agreed that the annotation was very useful. The text and sequence tools were seen even more favorable, with more than two out of three agreeing that they are very useful. While some could argue that this is to be expected, since text is the basis of communicating in most education, it also adds certainty that our proposed text system is at least adequate.

\subsection{Limitations}

While this study provides valuable insights into the utilization of annotations in virtual reality environments, it is crucial to acknowledge some considerations that may have influenced the outcomes.
One aspect is the measurement of durations during the study. The method employed involved capturing the start and end times of events, which, in some cases, included passive phases. For instance, reported times for tool usage on the touchscreen encompassed periods when participants had the tool active without utilizing it. This method might have introduced variations in the recorded durations. 
An additional challenge surfaced regarding time constraints. Although the study was scheduled for two hours per participant, some individuals required more time. Consequently, this occasionally led to brief interruptions when accommodating the next participant. While this dynamic could potentially impact the continuity of the study, it underscores the engagement and thoroughness of participants.
A potential factor influencing participant experiences was the number and length of forms they were required to complete. Feedback from some participants indicated that the forms were perceived as lengthy, potentially contributing to moments of fatigue. It is essential to recognize that participant fatigue may have influenced the quality of responses in later stages of the study.

%% file: sections/07_conclusion.tex
\section{Conclusion}
In conclusion, our quantitative user study, involving 24 participants, assessed the impact of different annotations on learning and teaching within touchscreen and VR environments. Through this investigation, we aimed to explore the usability of annotations and interfaces designed as outlined in this paper.

The study revealed valuable insights into the usability of annotations in mixed environments. Participants, tasked with annotating scenes using the touchscreen interface, demonstrated the ability to transfer medical textbook material. While acknowledging the potential for refinement in the touch interface, the study affirmed its utility and showcased a broad range of possible applications. As learners immersed in the VR environment, participants expressed a strong affinity for the appeal of our VR setting. Their feedback provided valuable suggestions for diverse areas of application within the virtual realm. This positive reception underscores the potential of our work in creating engaging and effective learning experiences in virtual reality mixed with touchscreen educators.

Our work contributes to the establishment of a foundational framework for the utilization of annotations in both learning and teaching contexts within mixed environments that integrate touchscreens and virtual reality technology. By bridging the gap between these two modalities, our research opens avenues for innovative educational applications. While our study demonstrated promising results, we acknowledge certain limitations, such as the need for further refinement of the touch interface. Future research could explore enhancements to address these limitations and delve into specific areas suggested by participants for VR applications. 

In summary, this study advances our understanding of the practical implementation of annotations in mixed reality educational settings. The positive feedback from participants and the demonstrated usability of our proposed interface underscore the potential for enriching learning and teaching experiences through the synergy of touchscreens and virtual reality.